\documentclass[a4paper,11pt]{article}
\usepackage{jheppub} 
\usepackage{lineno}

\usepackage{slashed,nicefrac}

\usepackage{mathtools}

\usepackage[T1]{fontenc}

\usepackage{amssymb,amsmath}

\usepackage{xcolor,bbold,comment}

\usepackage{amsthm}

\arxivnumber{2407.12717} 

\title{\boldmath Consistency conditions for O-plane unsmearing from second-order perturbation theory}

\author{Maxim Emelin}
\affiliation{{\it Dipartimento di Fisica e Astronomia ``Galileo Galilei''\\
			Universit\`a di Padova, Via Marzolo 8, 35131 Padova, Italy}}
\affiliation{{\it  INFN, Sezione di Padova \\
		Via Marzolo 8, 35131 Padova, Italy}}

\emailAdd{maxim.emelin@pd.infn.it}

\abstract{Scale-separated AdS compactifications of string theory can be constructed at the two-derivative supergravity level in the presence of smeared orientifold planes. The unsmearing corrections are known to leading order in the large volume, weak coupling limit. However, first-order perturbative approximations of non-linear problems can often produce spurious solutions, which are only weeded out by additional consistency conditions imposed at higher orders. In this work, we revisit the unsmearing procedure and present consistency conditions obtained from the second order warp factor and dilaton equations. This requires proper treatment of the near-source singularities. The resulting conditions appear as integral constraints on various non-linear combinations of the first order corrections, which we argue can generally be satisfied by appropriate choice of integration constants of the leading-order solutions. This provides a non-trivial consistency check for the perturbative unsmearing procedure and supports the existence of scale-separated AdS vacua in string theory.}

\begin{document}
\maketitle
\flushbottom

\section{Introduction}

In order to produce a genuinely lower-dimensional effective description of a string compactification, one needs to be able to separate the compactification scale from other physical scales of interest. For cosmological purposes, a minimal requirement is a separation, parametric or otherwise, between the compactification and the cosmological scales. 

Given the various challenges in constructing the more cosmologically relevant de Sitter vacua \cite{Danielsson:2018ztv}, the more readily available compactifications to anti-de Sitter space might serve as an interesting alternative setting to study the possibility of scale-separation. Of course this assumes that the issues of scale separation and positive spacetime curvature are independent. 

Even in AdS, the question of scale-separation in string compactifications involves the interplay of many considerations (see \cite{Coudarchet:2023mfs} for a recent review). There are various arguments placing limitations on scale-separation within the paradigm of the swampland program \cite{Palti:2019pca, vanBeest:2021lhn}. Some arguments combine the Weak Gravity Conjecture \cite{Arkani-Hamed:2006emk} with extended supersymmetry to rule out scale-separated vacua in extended supergravity theories \cite{Cribiori:2022trc}. Other restrictions come from an application of the Swampland Distance Conjecture \cite{Ooguri:2006in} to AdS vacua. This results in the strong AdS Distance conjecture \cite{Lust:2019zwm}, which forbids scale-separation outright, with refinements proposed in \cite{Buratti:2020kda, Moritz:2017xto, Blumenhagen:2019vgj}. Other arguments are based on general features of the higher dimensional supergravity equations of motion \cite{Gautason:2015tig}, similar in spirit to arguments restricting de Sitter vacua \cite{Gibbons:2003gb, Maldacena:2000mw}. While there is no all-encompassing no-go theorem, these and other considerations restrict the corners of string theory where scale-separated AdS vacua can be found.

At the same time, scale-separated AdS vacua appear among some of the earliest constructions of orientifold flux compactifications. The paradigmatic example is the DGKT construction in massive type IIA strings, \cite{DeWolfe:2005uu} and its various extensions \cite{Camara:2005dc, Narayan:2010em, Marchesano:2019hfb}. This construction produces a family of AdS$_4$ flux vacua, with fully stabilized moduli, that allows parametric separation between the AdS radius and the size of the internal manifold. The scale-separated limit is simultaneously a weak-coupling, large volume limit, where the whole solution can be described in terms of 10-dimensional massive type IIA supergravity with orientifold sources. Similar solutions exhibiting scale separation have also been constructed in other dimensions and duality frames \cite{Cribiori:2021djm, Farakos:2020phe, Farakos:2023wps, VanHemelryck:2022ynr, Carrasco:2023hta}.

These solutions either provide workarounds for some of the assumptions in the arguments against scale-separation, or for some of the stronger conjectures, appear to be explicit counterexamples. The big caveat to the DGKT-type constructions is that they are either derived within a lower-dimensional effective field theory approach, or equivalently using a smeared approximation for the O-planes \cite{Acharya:2006ne}.

The smearing of sources is a common approximation in string compactifications, generally interpreted as a truncation of the full 10-dimensional equations of motion with localized sources to the internal zero-modes of all the fields. Since the sources actually couple to every Kaluza-Klein mode, this truncation is not a consistent one. The hope, instead, is that the truncation to zero-modes, serves as a valid approximation to the low-energy effective theory one obtains by integrating out the Kaluza-Klein modes. From the 10-dimensional perspective this is equivalent to the smeared solution accurately capturing the behavior of the fields away from the sources.

In practice, the relationship between smeared solutions and those with localized sources is not always straightforward. In the simplest cases of parallel BPS sources, one can construct both smeared and localized solutions and find that they approach each other in the large volume limit \cite{Baines:2020dmu}. On the other hand, there are also examples where both smeared and localized solutions are available, but do not share the same properties \cite{Blaback:2011nz,Blaback:2011pn,Apruzzi:2015zna}. Non-BPS sources are also known to be problematic for smearing \cite{Blaback:2010sj}.

In the case of DGKT vacua, only the smeared solutions are known explicitly \cite{Acharya:2006ne}. The construction of a full solution with localized sources is complicated by the presence of multiple source orientations, including intersection points in some limits. While not necessarily a conceptual problem, intersecting sources greatly complicate the construction of explicit metrics, with simple choices of ansatz known to fail in certain cases \cite{Bardzell:2024anh}. Removing the intersections by deforming the manifold, when this is possible, moves the technical challenge to the background manifold itself.

An important step toward putting these vacua on a solid footing was the construction of leading order perturbative solutions approximating the unsmeared solution as a correction to the smeared one \cite{Junghans:2020acz, Marchesano:2020qvg}. The procedure has also been successfully applied to the lower-dimensional and massless type IIA cases \cite{Emelin:2022cac, Cribiori:2021djm}. The success of these calculations appears to suggest a one-to-one correspondence between smeared and unsmeared DGKT-type solutions.

In the present work we will investigate further the consistency of this perturbative unsmearing procedure at higher orders. Indeed, the leading order correction, while suggestive, is blind to many of the potentially problematic features of the configuration, such as O-plane intersections and non-linearity of the equations of motion more generally. Of course, even if the procedure is perfectly consistent, obtaining the full higher-order corrections will be increasingly challenging and we do not aim to do so here. Instead, we will argue that the second-order equations already contain certain constraints on the leading-order solution before one even begins solving it. It is these consistency conditions that we will be interested in.

In section \ref{branches} we motivate our investigation of the higher-order equations. We outline the  sorts of troubles one encounters in perturbative approaches to non-linear problems and in gravity theories on compact spaces more specifically. We will argue for the insufficiency of the first-order result as an indicator of the existence of unsmeared solution and the necessity of investigating the higher-order equations. In section \ref{order1} we review the leading order perturbative unsmearing procedure with focus on features relevant to the second-order equations. In section \ref{order2} we will write down the second order equations and derive integral constraints on the lower order solutions focusing on the warp factor and dilaton equations. The derivation of the constraint will require careful treatment of the singular near-source behavior of the leading order corrections. We also comment on other possible constraints as well as the effect of O-plane intersections, which appears quite benign as far as our calculations are concerned.

At the end of our investigation, we find a rather optimistic result. The near-source behavior of the first-order corrections sources precisely the correct near-source behavior of the second-order warp factor and dilaton. This in turn enables formulating an integral constraint for the non-singular parts of the leading order corrections, which boil down to equations for the integration constants of the first-order corrections, undetermined by the leading order equations.

\section{Spurious solutions in first-order perturbation theory} \label{branches}

When applying perturbation theory to non-linear problems, there is an interesting subtlety that one must be aware of. Since the first-order perturbative expansion represents a linearization of the problem, its solutions will generally obey a superposition principle. Most of the time this is not a problem, but simply a reflection of the fact that the space of deformations of the unperturbed solution forms a manifold, and solutions of the linearized equations thus live in its tangent space. A problem occurs when the unperturbed solution is not a regular point on the manifold, but lies at the intersection of distinct branches of deformations. As a simple algebraic example, consider the equation
\begin{equation}\begin{aligned}
x(y-x) = 0 \ .
\end{aligned}\end{equation}
For $y = y_0 \neq 0$, we have the solutions
\begin{equation}\begin{aligned}
x = 0 \ , \qquad x = y_0 \ ,
\end{aligned}\end{equation}
and perturbing $y = y_0 + \epsilon$ and expanding $x$ in a power series around either of the original solutions as
\begin{equation}\begin{aligned}
x = x_0 + \epsilon x_1 + \frac12 \epsilon^2 x_2 + ...  \ ,
\end{aligned}\end{equation}
we have either
\begin{equation}\begin{aligned}
x_0 = 0  \implies x_1 = 0  \qquad \text{or} \qquad x_0 = y_0 \implies x_1 = 1 \ ,
\end{aligned}\end{equation}
which are precisely in correspondence with the tangent vectors to the two branches of solutions. For $y_0=0$ however, the situation is different. Plugging in the $x$ expansion we have $x_0 = 0$ and the term linear in $\epsilon$ becomes
\begin{equation}\begin{aligned}
x_1 x_0 + x_0 (1-x_1) = x_0 = 0 \ ,
\end{aligned}\end{equation}
which is tautologically true and does not restrict $x_1$. 

As advertised, the problem is that the first order equation is a linearization of the original problem around the intersection of two branches of solutions and the tangent vectors to these branches span a whole vector space of solutions to the linearized equation, which are themselves not tangent to any branch of solutions. The situation is quickly remedied by going to the next order in perturbation theory. At order $\epsilon^2$ we have
\begin{equation}\begin{aligned}
\frac12 x_2(y_0-x_0) + x_1(1-x_1)-\frac12 x_0 x_2 = 0 \ ,
\end{aligned}\end{equation}
which for $x_0 = y_0 = 0$ gives
\begin{equation}\begin{aligned}
x_1(1-x_1) = 0 \ ,
\end{aligned}\end{equation}
once again reducing the set of values for $x_1$ to only those solutions that are tangent to the branches.

This extra hassle is due to the special nature of the point $x = y = 0$. Indeed settings with an enhanced degree of symmetry are particularly prone to producing spurious solutions at leading order in perturbation theory. 

In the context of general relativity and specifically linearized gravity, this problem is well known, and is sometimes referred to as linearization instability \cite{AltasKiraci:2018rvh, Altas:2019qcv}. In particular, it occurs when studying deformations of pure GR solutions on spatially compact manifolds with killing vectors \cite{Fischer,Deser:1973zzb}. A paradigmatic example is Einstein gravity on $\mathbb{R} \times T^3$. Obviously, the flat metric is a solution, differing from Minkowski space only by the periodic conditions on the spatial directions. To look for perturbative deformations, we can expand
\begin{equation}\begin{aligned}
g_{\mu\nu} = g^{(0)}_{\mu\nu} + \epsilon h^{(1)}_{\mu\nu} + \frac12 \epsilon^2 h^{(2)}_{\mu\nu} + ... \ ,
\end{aligned}\end{equation}
and at linear order in $\epsilon$ we obtain the familiar gravity wave solutions of linearized gravity, satisfying
\begin{equation}\begin{aligned}
\Box \tilde{h}_{\mu\nu}=0  \ , \qquad \tilde{h}_{\mu\nu} = h_{\mu\nu}^{(1)} - \frac12 g_{\mu\nu} h^{(1)\rho}_{~~~\rho} \ ,
\end{aligned}\end{equation}
with the only difference from the non-compact case being the momentum discretization due to the periodic boundary conditions.

At the next order, the non-linearities of the Einstein equations begin to show up and the equations take the schematic form
\begin{equation}\begin{aligned}
0 = \mathcal{A}(\nabla \nabla h^{(2)}) + \mathcal{B}( h^{(1)} \nabla^2 h^{(1)}) +  \mathcal{C}( \nabla h^{(1)} \nabla h^{(1)} ) \ ,
\end{aligned}\end{equation}
with $\mathcal{A},\mathcal{B},\mathcal{C}$ linear functions of the various index contractions within their arguments. Notably, the functions $\mathcal{B}$ and $\mathcal{C}$ are such that their sum is not a total derivative for all $h^{(1)}$. $\mathcal{A}$, on the other hand is a total derivative and thus integrates to zero over the compact directions. This imposes an integral constraint on $h^{(1)}$, which is not satisfied by all the solutions to the linearized equations. In particular, it is violated by a simple plane wave travelling around one of the compact directions.

This problem is easily avoided by relaxing the background ansatz and including an additional mode corresponding to the overall size of the torus. The physical interpretation is rather simple: the plane wave carries additional energy and momentum compared to the flat solution and its backreaction leads to a change of the torus size. Conversely, insisting on a constant size torus means that the total energy of excitations must vanish, a restriction that is not detectable at the linearized level, but does manifest at higher order. 

More recently, examples of this linearization instability involving AdS space, rather than compact spatial manifolds, as well as some matter-coupled gravity theories have also been studied \cite{Altas:2017fcp, Altas:2018dci}. In all these contexts, the conditions coming from second order perturbation theory take the form of integral constraints where a certain topological quantity must be canceled by the integral of non-linear combinations of the first-order perturbations.

Of course 10-dimensional supergravity with localized sources is a much richer system than the ones that have been studied, and we do not pursue a systematic investigation of linearization instability in this context. It is, however, rather notable that the smeared DGKT-type solutions seem to feature all the ingredients where the linearization instability shows up, i.e. AdS space and in particular its killing vectors, highly symmetric compact manifolds, including 3-tori transverse to the O-planes. On the other hand, one of the main properties of O-planes is that they circumvent no-go conditions derived in a very similar way, by integrating equations of motion over compact spaces. In that case, it is interesting to see how this manifests in the perturbative approach. 

For another perspective on why the perturbative solution may be misleading, recall that the smeared DGKT-type solutions also feature a non-standard scaling symmetry, 

\begin{equation}\begin{aligned}
\tau \sim n^{3/4} \ , \quad w \sim n^{3/4} \ , \quad F_p \sim n^{p/4}\ , \quad H_3 \sim n^0  \ ,
\end{aligned}\end{equation}
which is different from the natural scaling symmetry of the supergravity equations themselves. This allows us to make contact with the discussion of intersecting branches of solutions via the following heuristic argument. Consider the smeared, large volume, zero curvature limit of a collection of O-plane sources. In this limit, only the zero-modes of all the fields and the source density are non-vanishing. On the one hand, we know that at least in the case of parallel sources with flat worldvolumes one can unsmear the sources exactly. Thus switching on the non-zero-modes of the source density and the appropriate fields is at least sometimes a valid deformation away from this limit. On the other hand, the scale-separated limit of smeared DGKT-type compactifications also results, by construction, in a rapidly vanishing AdS curvature, resulting again in a configuration with large volume and smeared sources with vanishing worldvolume curvature. Thus the smeared DGKT-type solutions appear to form another family of deformations away from a similar limit\footnote{The limits are not strictly identical, because the AdS case necessarily involves multiple source orientations. For the purposes of this heuristic argument we regard this as an orthogonal issue.}, but keeping the non-zero-modes vanishing. The question of the existence of unsmeared DGKT-type solutions thus boils down to whether these two types of deformations from the large volume smeared flat source limit can be combined. 

The scenario is thus analogous to the discussion of the algebraic example earlier. If both deformations are valid individually, a first-order calculation around this limit will necessarily conclude that simultaneously switching on a scale-separated AdS curvature and the non-zero modes of fields and sources is legitimate, simply by linearity. Whether these deformations represent distinct incompatible branches or are subsets of a larger manifold of valid deformations can only be distinguished by going to higher order.

All this suggests that an investigation of the higher order terms in the unsmearing procedure is more than an exercise in increasing the precision of the solution. As we will see, the general feature of higher-order perturbation theory imposing integral constraints on lower-order perturbations holds true and is quite tractable in this context. We will find that in contrast to the pure gravity case discussed above, the leading order unsmeared solution passes these consistency checks in a way that involves the precise near-source singular behavior of the fields as well as the nature of the source singularity itself in a rather non-trivial fashion.

\section{Scale-separated AdS and leading order unsmearing} \label{order1}
In this section we review the relevant features of the DGKT setup and the unsmearing procedure at leading order, following the approach of \cite{Junghans:2020acz} . The setting is massive type IIA supergravity with negative tension sources (O-planes). The metric ansatz is
\begin{equation}\begin{aligned}\label{metric}
ds^2 = g_{ij} dx^i dx^j+ g_{mn} dy^m dy^n \ ,  \qquad  g_{ij} = w(y)^2 \hat{g}_{ij} \ ,
\end{aligned}\end{equation}
where $\hat{g}_{ij}$ is a unit radius AdS metric. The $y$-coordinates parametrize the internal space and have $\mathcal{O}(1)$ range or periodicity.
The NSNS flux $H_3$ and the RR fluxes $F_2, F_4$ and $F_6$ are all chosen to be internal. There is also a non-vanishing Romans mass $F_0$ and a dilaton $\tau = e^{-\phi}$. Finally, the O-plane source terms are described by projectors specifying their orientation, defined in the usual way by
\begin{equation}\begin{aligned}
\Pi_{mn}^{(i)} = g_{mn} - \sum_{k=1}^3 n_{m}^{(i)k} n_{n}^{(i)k} \ ,
\end{aligned}\end{equation}
where $i$ labels the various separate sources and $n^{(i)k}$ are vielbeins orthogonal to the $i$-th source. The transverse metric is then
\begin{equation}\begin{aligned}
g^{(i)\perp}_{mn} = g_{mn} - \Pi^{(i)}_{mn} = \sum_{k=1}^3 n_{m}^{(i)k} n_{n}^{(i)k} \ .
\end{aligned}\end{equation}
The charge densities for each source can be written as
\begin{equation}\begin{aligned}
\rho^{(i)} = \frac{\delta^{(3)}(y_{\perp}^{(i)})}{\sqrt{g^{(i)}_\perp}} \ ,
\end{aligned}\end{equation}
where $y_\perp^{(i)}$ are local coordinates orthogonal to the $i$-th source. 
We can also define transverse 3-forms
\begin{equation}\begin{aligned}
\rho^{(i)}_3 = \rho^{(i)} \bigwedge_{k=1}^3 n^{(i)k} \ , 
\end{aligned}\end{equation}
which will appear in the flux equations of motion. In local coordinates they can be written as
\begin{equation}\begin{aligned}
\rho^{(i)}_3 = \delta^{(3)}(y^{(i)}_\perp) dy_\perp^{(i)1} \wedge dy_{\perp}^{(i)2}  \wedge dy_\perp^{(i)3} \ .
\end{aligned}\end{equation}
\subsection{Equations of motion}
The equation of motion for the $F_2$ flux is
\begin{equation}\begin{aligned}\label{f2eq}
dF_2 &= F_0 H_3 - \sum_i \mu \rho^{(i)}_3 \ ,
\end{aligned}\end{equation}
where $\mu$ is related to the absolute value of the sources' charge/tension, which we may keep unspecified for our purposes. For the specific case of O6-planes, $\mu = 2$ when working on the covering space of the orientifold spacetime involution. The $H_3$ flux equation is
\begin{equation}\begin{aligned}\label{h3eq}
d (\tau^2 \star H_3) + \star F_2 F_0 + \star F_4 \wedge F_2 + \star F_6 \wedge F_4 = 0\ .
\end{aligned}\end{equation}
The remaining flux equations are
\begin{equation}\begin{aligned} \label{otherf}
dF_4 &= H_3 \wedge F_2 \ ,   \quad dF_6 = d\star F_6 = 0 \ , \quad dH_3 = 0 \\
d \star F_2 &+ H_3 \wedge \star F_4  = 0 \ , \quad 
d \star F_4 + H_3 \wedge \star F_6 = 0\ .
\end{aligned}\end{equation}
These are typically satisfied simply by taking co-closed fluxes and choosing legs such that the wedge products vanish. For simplicity we will assume that $F_6 = 0$, which will have no significant impact on any of the arguments.

For the metric and dilaton equations, it will be convenient to take the following combinations: 
\begin{equation}\begin{aligned}
\mathcal{E}_\tau &= \left( \tau \frac{\delta}{\delta \tau} + \frac12 g^{MN} \frac{\delta}{\delta g^{MN}} \right)  S_{IIA}  \\
\mathcal{E}_w &= -\frac14 \left(g^{ij} \frac{\delta}{\delta (g^{ij})} + \tau \frac{\delta}{\delta \tau} \right) S_{IIA} \\ 
\mathcal{E}_{mn} &= \left(\frac{\delta}{\delta g^{mn}} + \frac14 g_{mn} \tau \frac{\delta}{\delta \tau} \right) S_{IIA}  \ .
\end{aligned}\end{equation}
Writing these out explicitly gives
\begin{equation}\begin{aligned}
\mathcal{E}_{\tau} &= \tau \Box \tau + \partial_m \tau \partial^m \tau + 4 \frac{\tau}{w} \partial_m \tau \partial^m w - \sum_p \frac{p-5}{4} |F_p|^2 -\frac12 \tau^2 |H_3|^2 - \frac34 \tau \mu \sum_i \rho^{(i)}  \\
\mathcal{E}_{w} &= \tau^2 \frac{\Box w}{w} + 3 \tau^2 \frac{\partial_m w \partial^m w}{w^2} + 2 \frac{\tau}{w} \partial_m \tau \partial^m w - \frac14 \frac{\tau^2}{w^2}R^{(AdS_4)} - \frac14 \sum_p |F_p|^2 + \frac14 \tau \mu \sum_i \rho^{(i)} \\
\mathcal{E}_{mn} &=\tau^2 R_{mn} + 2 \partial_m \tau \partial_n \tau - 2 \tau \nabla_m \partial_n \tau - \frac{4 \tau^2}{w} \nabla_m \partial_n w \\
& \quad - \frac12 \tau^2 |H_3|_{mn}^2 - \frac12 \sum_p (|F_p|^2_{mn} - \frac12 g_{mn}|F_p|^2) - \frac12 \tau \mu \sum_i  (\Pi^{(i)}_{mn} - \frac12 g_{mn})  \rho^{(i)} \ ,
\end{aligned}\end{equation}
where $R^{(AdS_4)} = -12$. The first two equations only involve the Laplacian of either the dilaton or the warp factor, respectively, with other fields appearing with at most first derivatives. The form of the last equation eliminates the internal Ricci scalar. It also has the convenient feature in the $T^6/\mathbb{Z}_3^2$ case that the sum over images for the source terms cancels out exactly in the smeared limit, or integrates to zero in the localized case.

\subsection{Field expansions and leading order unsmearing equations}

We now review the calculation of the leading order unsmearing corrections in DGKT-type solutions. To this end, we expand all the dynamical fields in an asymptotic series of the form
\begin{equation}\begin{aligned}\label{expansion}
X = n^{\alpha_X} \sum_{k=0}^{\infty} \frac{1}{k!} X^{(k)} n^{-\beta_X k} \ ,
\end{aligned}\end{equation}
for each field $X$, where $\alpha_X$ and $\beta_X$ are suitably chosen constants and $X^{(k)}$ are dynamical.\footnote{The factor of $\frac{1}{k!}$ in the expansion is chosen to coincide with the conventions of the \emph{xPert} module of the \emph{xAct} package for Wolfram Mathematica \cite{Brizuela:2008ra}, which was used to carry out all the perturbative expansions.} The $\alpha_X$ are chosen to coincide with the scalings of the smeared DGKT-type solutions, i.e.
\begin{equation}\begin{aligned}
\alpha_{g_{mn}} = \frac12 \ , \quad \alpha_{w} = \frac34 \ , \quad \alpha_{\tau} = \frac34 \ , \quad
\alpha_{F_p} = \frac{p}{4} \ , \quad \alpha_{H_3} = 0 \ .
\end{aligned}\end{equation}
Note that the difference in scaling between $g_{mn}$ and $w^2$ is ultimately responsible for the scale-separation at large $n$. The choices of $\beta_X$ are in principle arbitrary, but in practice are inspired by the behavior of the equations at subleading order. 

Plugging in the expansion for each field, the leading order equations read
\begin{equation}\begin{aligned}\label{zerothFlux}
dF_2^{(0)} = 0  \ , \quad dF_{4}^{(0)} = 0 \ , \\
d ( \tau^{(0)2} \star H^{(0)}_3 ) = 0 \ , \quad d \star F_{p}^{(0)} = 0  ,
\end{aligned}\end{equation}
for the fluxes and
\begin{equation}\begin{aligned}\label{zeroth}
0 & = \tau^{(0)} \Box \tau^{(0)} + \partial_m \tau^{(0)} \partial^m \tau^{(0)} + 4 \frac{\tau^{(0)}}{w^{(0)}} \partial_m \tau^{(0)} \partial^m w^{(0)}  \\
0 & = \tau^{(0)2} \frac{\Box w^{(0)}}{w^{(0)}} + 3 \frac{\tau^{(0)2}}{w^{(0)2}} \partial_m w^{(0)} \partial^m w^{(0)} + 2 \frac{\tau^{(0)}}{w^{(0)}} \partial_m \tau^{(0)} \partial^m w^{(0)} \\
0 & = \tau^{(0)2} R^{(0)}_{mn} - 4 \tau^{(0)2} \frac{\nabla_m \partial_n w^{(0)}}{w^{(0)}} - 2 \tau^{(0)} \nabla_m \partial_n \tau^{(0)} + 2 \partial_m \tau^{(0)} \partial_n \tau^{(0)} \ ,
\end{aligned}\end{equation}
for the dilaton and metric equations. The covariant derivatives, index contraction and the Laplacian are all with respect to $g^{(0)}_{mn}$. These leading order equations imply that $\tau^{(0)}$ and $w^{(0)}$ are constant, all leading order fluxes are harmonic and the leading order internal metric is Ricci-flat. This is in line with the properties of the smeared solution. Note that the O-plane sources do not enter any of the leading order equations and tadpole cancellation is not imposed at this order.

We then choose the $\beta_X$ so that the terms that did not appear at leading order begin to contribute. This results in the values used in \cite{Junghans:2020acz}. 
\begin{equation}\begin{aligned}
\beta_{g_{mn}} = \beta_{w} = \beta_\tau = 1 \\
\beta_{H_3} = \beta_{F_p} = \frac12 \ .
\end{aligned}\end{equation}
Other options for the scalings of subleading terms were explored in \cite{Andriot:2023fss,Tringas:2023vzn}, which allow for deformations of the original solutions at orders before the source terms enter the equations. With our current choices, at next order, we obtain for the $F_2$ flux
\begin{equation}\begin{aligned}\label{firstFlux}
dF^{(1)}_2 = F_0 H^{(0)}_3 - \sum_i \mu \rho^{(i)}_3 \ .
\end{aligned}\end{equation}

It is at this order that the source terms first enter. In particular, integrating this equation over the internal manifold imposes tadpole cancellation, which sets $H_3^{(0)}$ to its value in the smeared solution. The term involving only the leading order fields can be replaced by a set of smeared source densities denoted $j^{(i)}$ resulting in

\begin{equation}\begin{aligned}
dF_2^{(1)} = - \mu \sum_i (\rho^{(i)}_3 - j^{(i)}_3 ) \ ,
\end{aligned}\end{equation}
with $j^{(i)}_3$ denoting the transverse 3-form for the $i$-th source, defined analogously to $\rho^{(i)}_3$.
The $H_3$ equation does not involve the sources and at this order takes the form
\begin{equation}\begin{aligned}
\tau^{(0)2} d \star H_3^{(1)} + F_0 \star F_2^{(0)} +  \star F_4^{(0)} \wedge F_2^{(0)} + \star F_6^{(0)} \wedge F_4^{(0)} = 0\ ,
\end{aligned}\end{equation}
The remaining flux equations take on a similar form. Integrating over the appropriate internal submanifold kills the term involving the first order corrections, and sets a condition on the zeroth order fields, obeyed by their smeared values. Setting them to those values implies, in turn, that the first order corrections are co-closed, and in fact harmonic, with respect to the zeroth order metric.

For the metric and dilaton equations we can do the same thing as for the $F_2$ flux. The terms involving only zeroth order fields get replaced by the sum of smeared source densities $j^{(i)}$. With this replacement, the next order equations become
\begin{equation}\begin{aligned}\label{dilmetSS}
0 &= \tau^{(0)} \Box \tau^{(1)}  - \frac34 \tau^{(0)} \mu \sum_i ( \rho^{(i)} - j^{(i)} )  \\
0 & = \frac{\tau^{(0)2}}{w^{(0)}} \Box w^{(1)}  + \frac14 \tau^{(0)} \mu \sum_i (\rho^{(i)} - j^{(i)}) \\
0 & = \tau^{(0)2} R^{(1)}_{mn} - 4 \frac{\tau^{(0)2}}{w^{(0)}} \nabla_m \partial_n w^{(1)} - 2 \tau^{(0)} \nabla_m \partial_n \tau^{(1)} \\ & \qquad - \frac12 \tau^{(0)} \mu \sum_i (\Pi^{(i)}_{mn} 
 - \frac12 g_{mn}) (\rho^{(i)} - j^{(i)} )  \ ,
\end{aligned}\end{equation}
where 
\begin{equation}\begin{aligned}
R^{(1)}_{mn} &\equiv \frac12 (\nabla_p \nabla_m g^{(1)~p}_{~~n} + \nabla_p \nabla_n g^{(1)~p}_{~~m} - \nabla^m \nabla^n g^{(1)~}_{~mn} - \Box g^{(1)}_{~mn}) \ .
\end{aligned}\end{equation}
In particular we can choose a single function $\Phi(y)$ such that
\begin{equation}\begin{aligned}\label{ratios}
4\frac{\Box w^{(1)}}{w^{(0)}}  = -\frac{4}{3} \frac{\Box \tau^{(1)}}{\tau^{(0)}} = -2 \left(\nabla^{m} \nabla^n g^{(1)}_{mn} - \Box g^{(1)m}_{~~~~m} \right) = \Box \Phi \ .
\end{aligned}\end{equation}
The equations are linear in the first order corrections and the sources, so the solution can be expressed as a sum of solutions for individual sources, i.e. we can write
\begin{equation}\begin{aligned}
\Phi &= \sum_i \Phi^{(i)} \\
\Box \Phi^{(i)} &= -  \frac{\mu}{\tau^{(0)}} (\rho^{(i)} - j^{(i)}) \ .
\end{aligned}\end{equation}
Similarly, the correction to the internal metric can be expressed as a sum over corrections sourced by each source individually. In the particular case of the $T^6/\mathbb{Z}_3 \times \mathbb{Z}_3$ orientifold, by choosing a traceless diagonal metric ansatz for each individual source, all the first order corrections can then be expressed in terms of the single function $\Phi$, up to integration constants \cite{Junghans:2020acz}.

\subsection{Near-source behavior of leading corrections} \label{near}
The equations \eqref{dilmetSS} fully determine the dominant behavior of the leading order corrections near the source loci, even without having to find a global solution. As this behavior will play a very important role in the second order calculation, we write it here explicitly. Near any one particular source, transverse spherical symmetry is restored to leading order in the transverse distance. We can use local coordinates $y^m$, such that the metric is diagonal at the source locus and $y^{1,2,3}$ are longitudinal to the source, while $y^{4,5,6}$ are transverse. We can also define a radial coordinate $r$ that measures proper transverse distance from the source, defined by
\begin{equation}\begin{aligned}
dr^2 = g^{(0)\perp}_{mn} dy^m dy^n = g^{(0)}_{44}dy^4 dy^4 + g^{(0)}_{55}dy^5 dy^5 + g^{(0)}_{66}dy^6 dy^6  \ .
\end{aligned}\end{equation}

For the warp factor and dilaton, the near-source behavior is fully specified by the behavior of $\Phi$
\begin{equation}\begin{aligned}
\Box \Phi = -\frac{\mu}{\tau^{(0)}} \frac{\delta(r) }{r^2} + ...  \\  
\Phi = \frac{\mu}{\tau^{(0)}} \  \  \frac{1}{r} + ... \ ,
\end{aligned}\end{equation}
where "..." denotes other non-divergent terms, coming from integration constants, deviations from the diagonal spherically symmetric form of the metric as we move away from the source, terms sourced by the zeroth order fields contained in $j^{(i)}$ and contributions from the other distant sources.
The expressions for the metric corrections depend on the choice of background coordinates as well as additional gauge freedom for the perturbation itself. Working in our locally rectangular coordinates we can write the corrections in the following simple form
\begin{equation}\begin{aligned}
g^{(1)\|}_{mn} = \frac\mu2 g^{(0)}_{mn} \frac1r + ... &= \frac12 g^{(0)}_{mn} \tau^{(0)} \Phi + ... \\
g^{(1)\perp}_{mn} = - \frac\mu2 g^{(0)}_{mn} \frac1r + ... &= -\frac12 g^{(0)}_{mn} \tau^{(0)} \Phi + ...  \ ,
\end{aligned}\end{equation}
which can be checked to satisfy the equations \eqref{dilmetSS} near the source. Note in particular that
\begin{equation}\begin{aligned}\label{traces}
g^{(1)m}_{~~m} = 0 + ...  \\
\tau^{(1)} - \frac12 \tau^{(0)} g^{(1)\perp m}_{~~~~m} = 0 + ... \ ,
\end{aligned}\end{equation}
in the near-source region. Finally, the $F_2^{(1)}$ flux also diverges at the source locus. Its expression is simplest in locally spherical transverse coordinates
\begin{equation}\begin{aligned}
F_2^{(1)} = - \mu \sin \theta d\theta \wedge d\phi + ...  \ ,
\end{aligned}\end{equation}
but can also be written in the rectangular coordinates in component form as
\begin{equation}\begin{aligned}
F_{mn} = - \mu \ \epsilon^{\perp}_{mnp} \frac{y^p}{r^{3/2}} + ... 
 \ , 
\end{aligned}\end{equation}
with $m,n,p$ restricted only to the transverse directions. In particular,
\begin{equation}\begin{aligned}
|F_2|^2 = \frac{\mu^2}{r^4} + ... = \tau^{(0)2} \partial_m \Phi^{(i)} \partial^m \Phi^{(i)} + ...  \ .
\end{aligned}\end{equation}

\section{Second order equations and integral constraints} \label{order2}

So far, things look promising. The smeared solution has the interpretation as the leading order term in an asymptotic expansion around a weakly-coupled, large-volume and scale-separated limit, for which one can reliably compute a first order correction by solving a Poisson equation on the internal manifold. The correction turns out to be small on most of the internal manifold, except at the O-plane locus, where the 2-derivative equations are not reliable in the first place. This seems to suggest a one-to-one correspondence between smeared and unsmeared solutions, with the smeared solution accurately capturing the dynamics away from localized sources. As discussed in section \ref{branches}, however, leading order perturbation theory around highly symmetric configurations may yield spurious solutions, whose inconsistency is only revealed at higher orders. With this in mind, we proceed to examine the second-order corrections to the equations of motion. 

\subsection{Second order warp factor equation
}
We start with the warp factor equation, which at second order can be written as the sum of three pieces
\begin{equation}\begin{aligned}\label{warp2}
0 &= \tau^{(0)2} \frac{\Box w^{(2)}}{w^{(0)}} + \mathcal{E}_{w,\text{sing.}} + \mathcal{E}_{w,\text{reg.}} \ ,
\end{aligned}\end{equation}
where
\begin{equation}\begin{aligned}\label{wsing}
\mathcal{E}_{w,\text{sing.}} &= -\frac12 |F_2^{(1)}|^2 - 2 \frac{\tau^{(0)2}}{w^{(0)2}} w^{(1)} \Box w^{(1)}+ 6 \frac{\tau^{(0)2}}{w^{(0)2}} \partial_m w^{(1)} \partial^m w^{(1)} \\ 
& \quad +4 \frac{\tau^{(0)}}{w^{(0)}} \tau^{(1)} \Box w^{(1)} + 4 \frac{\tau^{(0)}}{w^{(0)}} \partial_m \tau^{(1)} \partial^m w^{(1)} \\
&\quad  - 2 \frac{\tau^{(0)2}}{w^{(0)}} g^{(1)}_{mn} \nabla^m \partial^n w^{(1)} - 2 \frac{\tau^{(0)2}}{w^{(0)}}  \nabla^m g^{(1)}_{mn} \partial^n w^{(1)} \\
& \quad +\frac{\tau^{(0)2}}{w^{(0)}}\nabla_m g^{(1)n}_{~~n} \ \partial^m w^{(1)} + \frac{1}{2} \mu \tau^{(1)} \sum_i \rho^{(i)} - \frac14 \mu \tau^{(0)} \sum_i (g^{(1)(i)\perp})^m_{~m} \  \rho^{(i)} \ ,
\end{aligned}\end{equation}
and
\begin{equation}\begin{aligned}\label{wreg}
\mathcal{E}_{w,\text{reg.}} & = R^{(AdS_4)} \frac{\tau^{(0)2}}{w^{(0)2}} \left( \frac{w^{(1)}}{w^{(0)}} - \frac{\tau^{(1)}}{\tau^{(0)}} \right) \\
& \quad - \frac12 F^{(0)}_{mn} F^{(2)mn} - \frac12 g^{(1)mn} |F_2^{(0)}|^2_{mn} \\
& \quad - \frac{1}{48} F^{(0)}_{mnpq} F^{(2)mnpq} - \frac12 g^{(1)mn} |F_4^{(0)}|^2_{mn} - \frac12 |F_4^{(1)}|^2 \ .
\end{aligned}\end{equation}
Note that the terms in $\mathcal{E}_{w,\text{sing.}}$ have divergent integrals near any individual source (away from intersections). Meanwhile terms in $\mathcal{E}_{w,\text{reg.}}$ have at most a $\frac{1}{r}$ type of divergence near each source, which has a finite integral the neighborhood of the source.

To obtain a constraint on the first order corrections, we proceed similarly to the example of Einstein gravity on a torus from section \ref{branches}. The idea is to integrate equation \eqref{warp2} over the internal manifold. This integral must, of course, vanish. However, unlike the linearized gravity example, it's not obvious that this integration eliminates the $\Box w^{(2)}$ term, despite its appearance as a total derivative. Indeed, the equation \eqref{warp2} must be satisfied everywhere on the internal manifold, but some terms have singular near-source behavior. {\it A priori}, it may be the case that some of this divergence is cancelled by the $\Box w^{(2)}$ term. This would mean that the integral of $\Box w^{(2)}$ alone over the full manifold is ill-defined and integrating the full equation does not produce a constraint on the first-order corrections alone. Excising small neighborhoods around the source loci from the domain of integration doesn't help, since $w^{(2)}$ would then contribute to the integral via boundary terms at the excisions that wouldn't vanish even as the size of the excised region goes to zero.

To see whether this is indeed a problem, we examine $\mathcal{E}_{w,\text{sing.}}$ in the neighborhood of a point on a single source, away from any intersection points, and use the near-source expressions from section \ref{near}. Doing so results in
\begin{equation}\begin{aligned}\label{nearW}
\frac{1}{\tau^{(0)2} }\mathcal{E}_{w,\text{sing.}} = - \frac{5}{8} \partial_m \Phi^{(i)} \partial^m \Phi^{(i)} -\frac{5}{8} \Phi^{(i)} \Box \Phi^{(i)}  + ... \ ,
\end{aligned}\end{equation}
where, again the $...$ denotes terms that have at most $1/r$ divergence. The expression \eqref{nearW} appears to be a total derivative, but both terms are singular near the source. The first term has $r^{-4}$ behavior, and thus must be canceled by the $\Box w^{(2)}$ term in \eqref{warp2}. This results in
\begin{equation}\begin{aligned}\label{wbulk}
w^{(2)} = \mu^2 w^{(0)} \frac{5}{16 r^2} + ...  \ , \qquad r > 0 \ .
\end{aligned}\end{equation}
The second term is more peculiar. Formally it has the form
\begin{equation}\begin{aligned}
\frac{\delta(r)}{r^3} \ ,
\end{aligned}\end{equation}
which takes us outside the realm of integrable functions (and distributions as limits thereof) and into more exotic generalized functions. This is a common occurence in non-linear theories and gravity in particular. A natural intepretation of such terms is that they determine the boundary conditions on the boundary of a small excised ball around the singularity, in the limit of vanishing ball radius, in such a way that integration by parts works.\footnote{More rigorous treatments of such objects exist in the literature e.g. \cite{Steinbauer:2006qi} in the context of general relativity. It would be interesting to apply these methods here, but for our purposes this simple treatment will suffice.} Thus, $w^{(2)}$ is given by \eqref{wbulk} for $r>0$, but must also contain a singular generalized function piece at $r=0$, which specifically cancels the divergent part of its integral. Any remaining finite contributions to $\Box w^{(2)}$ form a regular total derivative whose integral is well defined over the entire compact manifold and therefore vanishes. This allows us to proceed with the argument in the style of section \ref{branches} and write down the following integral constraint on the first order corrections.
\begin{equation}\begin{aligned}
\int d^6 y \sqrt{g^{(0)}} \left( \mathcal{E}_{w,\text{sing.}} + \mathcal{E}_{w,\text{reg.}} \right) = 0  \ .
\end{aligned}\end{equation}
The divergent contributions of $\mathcal{E}_{w,\text{sing.}}$ cancel by the reasoning above, but it also contains finite pieces, whose integrals must cancel against that of $\mathcal{E}_{w,\text{reg.}}$ . 

In order to satisfy the constraint, recall that the first order equations of motion only determine $w^{(1)}$, $\tau^{(1)}$ and $g_{mn}^{(1)}$ up to integration constants. The integral constraint is thus simply a linear equation for these integration constants. Of course there is more than one such constraint. For example, the dilaton equation also produces an integral constraint, which we examine next.

Before proceeding to the dilaton equation, we make a few remarks:
\begin{itemize}
 \item 
The equality of the coefficients in equation \eqref{nearW} is not trivial since \eqref{wsing} is not a manifest total derivative. For example, in \eqref{wsing}, only the second and third lines take the form of total derivatives. The first line only does so because the near-source behavior of the flux has the appropriate coefficent compared to the gradient of the warp factor. This, in turn, is only true because the O-plane has equal charge and tension. The cancellation of the source term in the last line of \eqref{wsing} is also non-trivial and results from the identical near-source behavior of the dilaton and the trace of the transverse metric.

\item
The $-5/8$ coefficient in \eqref{nearW} is precisely the one required to reproduce the expansion of the warp factor for a spherically symmetric O6 solution with flat worldvolume
\begin{equation}\begin{aligned}
w = (c - \mu/r + \mathcal{O}(r^2))^{-1/4} \ ,
\end{aligned}\end{equation}
for large integration constant $c$. Note that in the massive type IIA case, this is not the same as a large $r$ limit, because of the $\mathcal{O}(r^2)$ terms, which limit the domain of validity of the solution at large distance. This does, however, coincide with the large-$r$ expansion of the massless O6 in flat space. 

It is curious that this information is reconstructed from the singular near-source parts of the perturbative solution, given that the perturbative solution itself is meant to be accurate far from the sources.

The agreement with the behavior of the massless O6-plane might appear to suggest that the near-source behavior of the full solution is different from the proposed numerical solutions of \cite{Saracco:2012wc} for a localized O6-plane in massive type IIA and more similar to those in the massless theory. However, although at this order the coefficients of the $1/r^2$ terms match up, we expect further $1/n$ corrections to appear to this term if the computation is pushed to higher order. Thus it is difficult to foresee what the final function will resum to and whether it may indicate instead a behavior more similar to the solutions of \cite{Saracco:2012wc}.

\item 
We have so far not discussed the effect of possible O-plane intersections. Indeed, terms in \eqref{wsing} also contain cross-terms of the form
\begin{equation}\begin{aligned}
\Phi^{(i)} \Box \Phi^{(j)}  \ , \qquad \partial_m \Phi^{(i)} \partial^m \Phi^{(j)}
\end{aligned}\end{equation}
with $i \neq j$, whose divergences at their intersection locus feature two different transverse ``radial" directions. One may wonder if these divergences might cause additional problems. 

The $\Phi^{(i)} \Box \Phi^{(j)}$ terms, take the form
\begin{equation}\begin{aligned}
\frac{1}{r_1} \frac{\delta(r_2)}{r_2^2} \ .
\end{aligned}\end{equation}
 For an intersection such that no transverse directions along the internal space are shared, performing the integral over the directions transverse to the delta function we are left with a remaining integrand proportional to $1/r_1$ with the coefficient fixed by the intersection angles. Integrated over the remaining directions, this only gives a finite contribution. The same is true for the $\partial_m \Phi^{(i)} \partial^m \Phi^{(j)}$ terms, which are proportional to 
 \begin{equation}\begin{aligned}
\frac{1}{r_1^2}\frac{1}{r_2^2} .
 \end{aligned}\end{equation}
For a pair of sources with no common transverse directions, each factor is integrated over a separate subspace and gives a finite result.

Finally, using the explicit expressions from section \ref{near}, one can directly check that
\begin{equation}\begin{aligned}
\frac{1}{2} F^{(i)}_{mn} F^{(j)mn} = \tau^{(0)2} \partial_m \Phi^{(i)} \partial^m \Phi^{(j)}  \ ,
\end{aligned}\end{equation}
which precisely completes the near-intersection behavior of the first line of \eqref{wsing} to a total derivative form. Thus these types of intersections do not make any additional singular contributions to the integral constraints. At most, they give a finite contribution to the integration constants in the first-order corrections. 

Note that the necessary cancellations between the flux and other terms, necessary for the near-source behavior to take on a ``total derivative" form, would not hold true for an intersection with a shared transverse direction. Fortunately, such O-plane intersections in the DGKT construction result from an inconsistent choice of spacetime involution, as explained in \cite{Junghans:2023yue}. On the other hand, the allowed type of intersection features no common transverse directions and can be eliminated by blow-ups, in which case no extra divergent contributions to the integral constraint are possible. It is thus perhaps not that surprising that they are also absent in the intersecting limit.

\item 
Our analysis is restricted to the two-derivative equations of motion. As emphasized in \cite{Andriot:2023fss}, higher derivative terms begin to contribute at the same order as the second order perturbative corrections. One may therefore ask what their effect on our analysis might be. Since these terms involve higher derivatives of the first-order corrections, their near-source behavior will be more singular than any of the two-derivative terms. 

From here there are several possibilities. The best case scenario is that the higher-derivative terms appear with the appropriate coefficients so that their singular parts once again assemble into a ``total derivative"-like structure involving a divergent function in the neighborhood of the O-plane locus and a generalized function at the locus itself such that the total integral is finite. The effect of the higher derivative terms would then be to correct the consistency conditions through finite contributions, which are highly suppressed away from the O-plane locus. 

A more puzzling alternative would be that the higher derivative terms introduce mismatched singular contributions, invalidating integration by parts when integrating over the manifold and forbidding the formulation of integral constraints altogether. This would also complicate the interpretation of the generalized function terms localized at the O-plane locus, since they would no longer be equivalent to setting boundary conditions for small excisions. 

Although the first option would certainly be more appealing for our analysis, without knowing all the higher derivative corrections that contribute at a given order, it is impossible to say which option plays out. That said, it is still a relevant question whether two-derivative solutions exist to begin with, before worrying how they might be corrected or invalidated by higher-derivative corrections. The perturbative unsmearing procedure promises a systematic approximation scheme to a two-derivative solution. Our present analysis suggests that this procedure remains consistent at higher orders when applied to the second-order equations of motion.

\end{itemize}

\subsection{Second order dilaton equation, other constraints and overall consistency}

The dilaton equation paints a very similar picture to the warp factor equation. At second order the equation is
\begin{equation}\begin{aligned}\label{dil2}
0 &= \tau^{(0)} \Box \tau^{(2)}  + \mathcal{E}_{\tau,\text{sing.}} + \mathcal{E}_{\tau,\text{reg.}} \ ,
\end{aligned}\end{equation}
with
\begin{equation}\begin{aligned}\label{dilsing}
\mathcal{E}_{\tau,\text{sing.}} &= \frac32 |F_2^{(1)}|^2 + 8 \frac{\tau^{(0)}}{w^{(0)}} \partial_m \tau^{(1)} \partial^m w^{(1)} \\
& \quad + 2 \tau^{(1)} \Box \tau^{(1)} + 2 \partial_m \tau^{(1)} \partial^m \tau^{(1)} - 2 \tau^{(0)} g^{(1)}_{mn} \nabla^m \partial^n \tau^{(1)} - 2 \tau^{(0)} \nabla^m g^{(1)}_{mn}  \partial^n \tau^{(1)}  
\\
& \quad +\tau^{(0)} \nabla_m g^{(1)n}_{~~~n} \nabla^m \tau^{(1)} - \frac32 \mu \tau^{(1)} \sum_i \rho^{(i)} + \frac34 \mu \tau^{(0)} \sum_i (g^{(1)(i)\perp})^m_{~m} \  \rho^{(i)} \ ,
\end{aligned}\end{equation}
and
\begin{equation}\begin{aligned}\label{dilreg}
\mathcal{E}_{\tau,\text{reg.}} & = - \frac34 F^{(0)}_{mn} F^{(2)mn} - \frac32 g^{(1)mn} |F_2^{(0)}|^2_{mn} \\
& \quad + \frac{1}{48} F^{(0)}_{mnpq} F^{(2)mnpq} + \frac12 g^{(1)mn} |F_4^{(0)}|^2_{mn} + \frac12 |F_4^{(1)}|^2 \\
& \quad -\tau^{(0)}\tau^{(1)} |H_3^{(0)}|^2 - \frac{1}{12} \tau^{(0)2} H^{(0)}_{mnp} H^{(2)mnp} - \frac12 \tau^{(0)2} |H_3^{(1)}|^2 - \frac12 \tau^{(0)2} g^{(1)mn} |H_3^{(0)}|^2_{mn} \ .
\end{aligned}\end{equation}
Substituting the near-source expressions into $\mathcal{E}_{\tau, \text{sing.}}$ gives
\begin{equation}\begin{aligned}\label{neartau}
\frac{1}{\tau^{(0)2} } \mathcal{E}_{\tau, \text{sing.}} =  \frac{3}{8} \partial_m \Phi^{(i)} \partial^m \Phi^{(i)} + \frac{3}{8} \Phi^{(i)} \Box \Phi^{(i)}  + ...  \ .
\end{aligned}\end{equation}
The coefficients on the terms match once again and result in
\begin{equation}\begin{aligned}
\tau^{(2)} = -\mu^2 \frac{3}{16 r^2} + ... \ , \qquad r > 0  \ .
\end{aligned}\end{equation}
This is precisely the $1/r^2$ term in the large-$c$ expansion of
\begin{equation}\begin{aligned}
(c - \mu/r + \mathcal{O}(r^2))^{3/4} \ .
\end{aligned}\end{equation}

As before, in order to solve the equation on the full manifold, $\Box \tau^{(2)}$ must be a generalized function with a  $\delta(r)/r^3$ type singularity at the source locus such that its integral vanishes. This enables the integral constraint on the first order corrections
\begin{equation}\begin{aligned}
\int d^6 y \sqrt{g^{(0)}} \left( \mathcal{E}_{\tau,\text{sing.}} + \mathcal{E}_{\tau,\text{reg.}} \right) = 0 \ ,
\end{aligned}\end{equation}
which can be accommodated through suitable choice of integration constant for $\tau^{(1)}$ and $g^{(1)}_{mn}$. Note that $w^{(1)}$ does not appear without derivatives, so the two integral constraints we have seen fix independent combinations of the integration constants and are thus not mutually exclusive.

Finally, we might consider other constraints coming from the internal metric equation. One obvious and universally present constraint arises from the trace of the internal metric equation. The second-order expression is rather cumbersome, due to the multitude of possible index contractions, so we simply quote the results. 

The final expression for the singular terms in the near-source limit is
\begin{equation}\begin{aligned}
\frac12 \tau^{(0)2} \nabla^m \nabla^n g^{(2)}_{mn} - \frac12 \tau^{(0)2} \Box g^{(2)m}_{~~~m} & - \tau^{(0)} \Box \tau^{(2)} - 2 \frac{\tau^{(0)2}}{w^{(0)}} \Box w^{(2)} \\
&= -\tau^{(0)2} \left( \frac{21}{8} \partial_m \Phi \partial^m \Phi  + \frac{21}{8} \Phi \Box \Phi \right) + ... \ .
\end{aligned}\end{equation}
Once again, the coefficients on the right hand side match, enabling a proper integral constraint. The value of the coefficient can be checked to be consistent with the large $c$ expansion for 
\begin{equation}\begin{aligned}
g_{mn} = (c - \mu/r + \mathcal{O}(r^2))^{\pm 1/2} \ ,
\end{aligned}\end{equation}
with $+$ in the exponent for the transverse directions and $-$ for the longitudinal ones. 

Upon examining the terms containing possible integration constants from the first-order corrections, we find yet another combination of $w^{(1)}$ , $\tau^{(1)}$ and $g^{(1)}_{mn}$, different from those of the warp factor and dilaton equation, meaning that this constraint is compatible with the others.

The three constraints we have investigated will be generically present in any compactification with localized sources that admits a smeared approximation. Additional constraints are possible if the internal space contains killing vectors. In that case we expect that they fix additional integration constants relating to the volumes of internal cycles etc. In that case, all the consistency constraints would need to be compatible, which we expect to generically be the case.

The mutual consistency of all the constraints is perhaps not surprising. The entire point of DGKT solutions is that they fully stabilize the moduli. The integration constants thus simply correspond to shifts in these stabilized moduli due to the unsmearing corrections. It might be interesting to compare their values to the shifts calculated from corrections to the lower-dimensional effective potential in specific examples.

\section{Conclusion}

Confirming or disproving the consistency of DGKT-type scale-separated AdS compactifications with fully localized sources would be an important step in understanding questions of scale-separation in string theory. In this work, we have examined the perturbative O-plane unsmearing procedure for DGKT-type vacua at second order in the metric and dilaton perturbations. At this order the non-linearities of the supergravity equations of motion first manifest themselves. We investigated the possibility of additional non-linear constraints or inconsistencies, similar to linearization instabilities in general relativity, arising from integrating the second-order equations. We found that the singular near-source behavior of the first-order corrections leads to a consistent near-source behavior of the second-order corrections, provided one treats them as generalized functions, with an appropriate type of singular contribution on the source locus. This in turn allows for a consistent derivation of integral constraints for the first-order corrections. The system has enough degrees of freedom that these constraints do not lead to inconsistencies, but rather serve to fix the integration constants of the first-order corrections. We also found that O-plane intersections of the sort that can occur in DGKT-type constructions appear to have very benign behavior, at least at this order.

Overall our results indicate that the DGKT and similar constructions, as well as the perturbative unsmearing procedure more generally pass a rather non-trivial consistency check, supporting the plausibility of scale-separated AdS vacua in string theory.

\end{document}